\documentclass[lettersize,journal]{IEEEtran}

\usepackage[normalem]{ulem}
\usepackage{amsmath,amssymb,amsfonts}
\usepackage{caption}
\usepackage{subcaption}
\usepackage{algorithmic}
\usepackage{graphicx}
\usepackage{textcomp}
\usepackage{xcolor}
\usepackage{mwe}

\usepackage{booktabs}
\usepackage{siunitx}
\usepackage{longtable}
\usepackage{array}
\usepackage[symbol]{footmisc}

\begin{document}
\let\WriteBookmarks\relax
\def\floatpagepagefraction{1}
\def\textpagefraction{.001}



\title {Multivariate Physics-Informed Convolutional Autoencoder for Anomaly Detection in Power Distribution Systems with High Penetration of DERs}  

\author{{Mehdi~Jabbari Zideh and Sarika Khushalani Solanki}


\thanks{Authors are with the Lane Department of Computer Science and Electrical Engineering, West Virginia University, Morgantown, WV 26506, USA (e-mail: mehdijabbari@ieee.org, skhushalanisolanki@mail.wvu.edu)}

\thanks{\emph{Corresponding author: Mehdi Jabbari Zideh.}}}

\maketitle
\begin{abstract}
Despite the relentless progress of deep learning models in analyzing the system conditions under cyber-physical events, their abilities are limited in the power system domain due to data availability issues, cost of data acquisition, and lack of interpretation and extrapolation for the data beyond the training windows. In addition, the integration of distributed energy resources (DERs) such as wind and solar generations increases the complexities and nonlinear nature of power systems. Therefore, an interpretable and reliable methodology is of utmost need to increase the confidence of power system operators and their situational awareness for making reliable decisions. This has led to the development of physics-informed neural network (PINN) models as more interpretable, trustworthy, and robust models where the underlying principled laws are integrated into the training process of neural network models to achieve improved performance. This paper proposes a multivariate physics-informed convolutional autoencoder (PIConvAE) model to detect cyber anomalies in power distribution systems with unbalanced configurations and high penetration of DERs. The physical laws are integrated through a customized loss function that embeds the underlying Kirchhoff's circuit laws into the training process of the autoencoder. The performance of the multivariate PIConvAE model is evaluated on two unbalanced power distribution grids, IEEE 123-bus system and a real-world feeder in Riverside, CA. The results show the exceptional performance of the proposed method in detecting various cyber anomalies in both systems. In addition, the model's effectiveness is evaluated in data scarcity scenarios with different training data ratios. Finally, the model's performance is compared with existing machine learning models where the PIConvAE model surpasses other models with considerably higher detection metrics.

\end{abstract}




\maketitle

\section{{Introduction}}\label{Introduction}
\IEEEPARstart{T}{he} increasing integration of distributed energy resources (DERs) such as solar and wind power units adds to the nonlinear nature of power systems \cite{2024tiwaripf, 2021deyhimi, nazaralizadeh2024battery, nematirad2024novel}. The system operators require more robust and trustworthy algorithms to make confident decisions for the reliable and efficient operation of power grids. The ultimate model should comply with different operational conditions of over-changing power systems while ensuring that the predicted solutions remain consistent with the system's underlying principles even during the system's transitions from one state to another.

Despite recent progress in data analytics and deep learning (DL), modeling and analyzing nonlinear multiscale systems have remained a challenging task. This arises because the collected data is sparse, noisy, and incomplete in the complex physical and engineering systems and the data acquisition process is prohibitively expensive \cite{karniadakis2021physics}. As a result, the decision-makers face severe challenges in making confident decisions with partial information. In such circumstances, purely data-driven models suffer from poor generalization and lack of extrapolation for the data beyond the training window as they are unable to interpret the underlying physical principles and integrate them into their learning process \cite{karniadakis2021physics, zideh2023physics}. The highly frequent sparse and incomplete data requires that computational tools and DL algorithms penalize physically inconsistent solutions to provide robust predictions and guarantee the convergence of training procedure \cite{lai2024physics}. Therefore, a new class of deep learning algorithms was introduced to overcome the main challenges and limitations of purely data-driven DL models.

Physics-informed neural networks (PINNs) as promising solutions have been exploited to integrate the principled physical laws and theoretical constraints into DL training mechanisms to present plausible predictions even under data scarcity and partial information \cite{zideh2023physics, raissi2019physics, huang2022applications, cuomo2022scientific, di2022physically}. Constraining the space of admissible solutions by embedding the fundamental physical laws steers the training phase of DL models toward more interpretable predictions and provides improved performance in uncertain nonlinear dynamical systems. PINNs have been applied to solve many scientific problems from computational fluid dynamics \cite{jin2021nsfnets, cai2021physics} and power and energy systems dynamics and steady-state \cite{liu2024hybrid, kuang2024physics} to health \cite{kissas2020machine} and systems safety \cite{xu2023physics}.

In recent years, the landscape of cyber-physical anomaly analysis has shifted from purely data-driven models to physics-informed learning algorithms. Although traditional DL models have been applied to detect and identify anomalous conditions, their solutions may not be trustworthy when the system's conditions vary unpredictably \cite{mestav2022deep, tabassum2024cyber, 2023mehrzadreview, 2023sakhnini}. These models heavily rely on the vast amount of data and their performance degrades when data scarcity issues come to play. On the other hand, physics-informed learning models have been developed to enhance the operator's situational awareness and ensure that the predicted solutions of the learning models follow the physical laws. 

Given the challenges of data-driven models, researchers have increasingly employed PINN models for cyber-physical anomaly detection in power systems. These models encompass various techniques to integrate physical principles into the loss function of neural networks, modifying the architecture of data-driven models, and fusing physics-based and data-driven models. Authors in \cite{tong2021detection} developed a graph convolutional neural network (GCN) to detect transient faults in power transmission lines. The proposed method utilizes spatial information such as adjacency matrix and bus voltage signals to extract features. The work in \cite{li2023graph} employed a graph-based neural network (GNN) model that extracts the power grid topology changes and operational data to detect false data injection attacks (FDIAs). In Ref \cite{boyaci2021graph}, a real-time detector of FDIAs based on the physical connection of power grid buses and the correlations of the system measurements are proposed. Although these research studies showed highly accurate results in detecting cyber-physical anomalies, as they integrate spatial knowledge not domain-specific physical laws, the provided predictions may be physically inconsistent. 
In \cite{lakshminarayana2022data}, the researchers proposed a physics-informed loss function strategy to embed dynamical relationships of the Swing equations to detect load-altering attacks. The proposed strategy was employed to infer the attack parameters providing insights into the attack presence and its location. A generative-adversarial network (GAN) based strategy was proposed in \cite{zheng2021generative} to reconstruct phasor measurement unit (PMU) data to generate high-resolution data for event detection and classification. The algebraic equations of the generator buses were integrated into the training process of GANs to ensure that solutions follow the system's physical laws. The work in \cite{xu2022blending} fused the physics-based information of a model-based detector with the outputs of an LSTM-autoencoder deep learning model to identify FDIAs and recover system measurements. The physics-informed strategies in the mentioned literature were applied to the systems with balanced configurations while their efficacy for unbalanced systems was not investigated.
In \cite{Dwivedi2023dynamopmu}, authors developed a data-driven dynamics-based approach for anomaly detection that leverages the nonlinear dynamical behavior of distribution systems to perform the singular vector decomposition of the Hankel matrix alternative view of Koopman (HAVOK). The work in \cite{2021liphysics} integrates the physics-based relationship of voltage and current trajectories into the loss function of an autoencoder (AE) model through a rotated ellipse equation to detect high-impedance faults. Although these research studies successfully embedded physical knowledge into data-driven models, they were not tested for large-scale test cases.

To address the gaps in the existing literature, this paper proposes a multivariate physics-informed convolutional autoencoder (PIConvAE) for cyber anomaly detection in unbalanced power distribution grids. The proposed unsupervised model embeds the fundamental Kirchhoff's circuit laws into the loss function of the autoencoder to effectively detect various cyber anomalies without the need for labeled data \cite{zideh2023piconvae}.  
The main contributions of this work are summarized as follow:
\begin{itemize}
\item{The underlying physical principles of power systems are integrated through Kirchhof's circuit laws into the loss function of a multivariate autoencoder to provide highly accurate detection results for cyber anomaly detection in power distribution grids with high penetration of DERs.}
\item{The performance of the proposed multivariate PIConvAE model is evaluated through simulations on the modified 123-bus system and a $\mu$PMU data of a real-world feeder in Riverside, CA (RCA). The real-world data of load patterns and DERs including wind speed, solar irradiance, and ambient temperature are applied to the 123-bus system  to simulate the system with real-world scenarios. The selection of these systems is to show how the proposed model performs in real-world cases and large-scale systems.}
\item{The anomaly detection performance of the proposed model is assessed through training it on datasets that represent only 50\%, 30\%, and 10\% of the original dataset to show the efficacy of the physics-informed model for scenarios of data scarcity where the cost of extensive data acquisition is expensive.}
\item{Comparative case studies are performed to compare the detection performance of the proposed PIConvAE model with the existing machine learning methods for cyber anomaly detection demonstrating the superior performance of the physics-driven model for both distribution systems.}
\end{itemize}

The remainder of this paper is organized as follows. Section \ref{cyber_Development} introduces different cyber anomaly functions. Section \ref{PIConvAE} presents the problem statement, underlying physical laws, and the development of the multivariate PIConvAE framework. The experimental results including system and data description, data preprocessing, model configuration, performance metrics, and detection results are presented in Section \ref{Results}. Finally, the concluding remarks are presented in Section \ref{conclusion}.

\section{Cyber Anomaly Models}\label{cyber_Development}
In this section, different types of attack functions including two special types of FDI attacks \cite{zideh2024unsupervised, musleh2023spatio}, ramp attack, replay attack, and denial-of-service attack \cite{hasnat2022graph} are introduced. These functions can be applied to time-varying power system signals such as voltage and current phasors, and active and reactive power measurements. Let us consider time-series measurements of $\mu$PMU devices synchronized in time. The corrupted data under cyber anomalies in the time interval ($t_{start}$, $t_{end}$) can be defined by a generalized cyber anomaly function as follows:
\begin{equation}\label{general_anomaly}
x^{attacked}=f(t) \quad {for} \quad t_{start}\leq t\leq t_{end}
\end{equation}
where $x^{attacked}$ can represent any of manipulated voltage and current magnitudes, voltage and current phase angles, and active and reactive powers. The function $f(t)$ is defined according to different cyber anomaly shapes and characteristics to model their effects on the power system time-series measurements. In the following, we describe each attack type.

\subsection{False Data Injection Attacks (FDIAs)}\label{FDIA}
FDIAs are sophisticated cyber anomalies launched by attackers to deceive the bad data detector (BDD) systems while meeting the state estimation (SE) criteria. In the power system SE, the relationship between power system measurements $z$ and states $s$ are represented through a nonlinear function $h$ as $z=h(s)$. The attackers' objective is to bypass the BDD mechanism by injecting false measurement $z^{att}(t)=f(t)$ to meet the residue of SE, $\|z^{att}-h(\hat{s})\|_2 \leq\tau$, where $\hat{s}$ is the estimated state. If the SE residue for a specific measurement is greater than the predefined threshold $\tau$, that measurement is flagged as an anomaly. The system state can be any of the power flow outputs including voltage and current phasors, and active and reactive powers.  In this work, the voltage magnitudes are considered as system states which can be obtained by installing power measurement devices (PMU or $\mu$PMU) or estimated with the available measurement devices throughout the network. In this section, two types of FDIAs are investigated:\\
\textit{\textbf{a) Combined (camouflage) attack}}: 
This type of FDIA is constructed by adding or deducting small values to/from real measurements that can bypass bad data detectors and cannot be easily detectable. We use the general model of combined attacks in \cite{zideh2024unsupervised} as $f(t)=z(t) +{(-1)^b \Delta z}$  
where $b \in \{0, 1\}$ with $b=0$ as additive attack and $b=1$ as deductive attack. The additive or deductive parts are selected such that the residual of the true measurements and the falsified data is within the detector threshold $\tau$.\\
\textit{\textbf{b) Stealth attack}}:
Stealth attack is a special type of FDIA where adversaries have the minimum information about system measurements to change the system observations over time. Here, the stealth attack model from \cite{musleh2023spatio} is employed which is represented as $f(t) = \alpha (t)(z(t)+\beta(t))$ where $\alpha(t)$ and $\beta(t)$ are time-varying attack multiplier and addend, respectively. These coefficients are chosen to be within the BDD threshold to satisfy the SE conditions and not be easily detectable.

\subsection{Replay Attack}\label{replay}
In this type of attack, the system measurements need to be tracked and recorded to be reused at different time intervals. The previously recorded measurements from measurement devices are inserted as the true values in the same or other measurement devices \cite{hasnat2022graph}. The replay attack is at timestamp $t$ is modeled as $f(t) = z^{record}(t_p)$, \  $t_p<t_{start}$. Replay attacks are challenging to detect because the recorded measurements represent the system's dynamics. The magnitudes of the replay attacks may vary from slight changes to large values depending on the compromised meter and the time of the recorded measurements.

\subsection{Ramp Attack}\label{ramp}
Ramp attack involves gradual changes in the system measurements to deceive the system operator. The detection of ramp attacks can be very challenging for the detection mechanisms and machine learning algorithms since there would be no abrupt change at the onset of the attack. The ramp attack is modeled as $f(t) = z(t_{start}) + m \times {(t - t_{start})} + q(t)$, where $m$ is the slope of the corrupted measurement and can be positive or negative and $q(t)$ represents the Gaussian noise. In this work, the ramp attacks are generated with slopes of 0.25 and -0.25.

\subsection{Denial-of-service (DoS) Attack}\label{DoS}
In the literature, a DoS attack is defined as the suspension of reliable access to enough power through compromised data and misleading the system operators about the system state estimation \cite{huseinovic2020survey}. It prevents data packets from being transmitted through communication channels leaving the system monitoring with only noise in the duration of such attack \cite{huo2023distributed}. The detection of DoS attacks is comparatively easier than other attack types as there will be abrupt changes during the occurrence of such attacks. To make it challenging for the detection algorithms, similar to \cite{hasnat2022graph}, the DoS attacks are considered as constant values during the attack time interval. It is assumed that the attacker prevents updating the time-series measurements of the measurement devices at the targeted locations meaning that the last recorded measurement before the onset of the attack is reported throughout the DoS attack duration. Specifically, the DoS attack function is expressed as $f(t)=z(t_{start})+q(t)$ for $t_{start}\leq t\leq t_{end}$, where $q(t)$ represents the Gaussian noise components.

\section{Physics-informed Learning Methodology}\label{PIConvAE}
This section outlines the main principles for constructing the physics-informed learning strategy for anomaly detection. In general, four paradigms have been introduced to integrate the underlying physical principles into neural network models \cite{zideh2023physics, huang2022applications}; 1) Physics-informed design of architecture, which refers to the embedding of the physical information into the architecture of NN models \cite{djeumou2022neural}. Graph neural networks are special classes of physics-informed design of architectures where the system's information is integrated into NN models through a set of nodes and edges \cite{zhou2020graph, ngo2024physics}. 2) Physics-informed loss function, where the physical principles are integrated into NN models as regularization terms to penalize the solutions that deviate from the physical laws \cite{raissi2019physics}. 3) Hybrid physics-based NN models, which fuse the white-box nature of physics-based simulations and black-box features of NN models to provide physically consistent predictions \cite{chen2020rotor}. 4) Physics-informed initialization, which is similar to transfer learning methods where the NN parameters are initialized through training with synthetic data and fine-tuned with fewer real-world observational data \cite{jia2021physics}. In this work, we exploit the underlying physical relationships between power system measurements by embedding them in the loss function of an autoencoder to reconstruct the input data accurately. The physics-based loss terms serve as regularization agents eliminating the need for other regularization techniques, i.e., L1 or L2 regularization.
\begin{figure*}
    \hspace{0mm}
    \centering
    \includegraphics[clip,trim=1.8cm 5.5cm 1.15cm 2.5cm, width=0.99\textwidth]{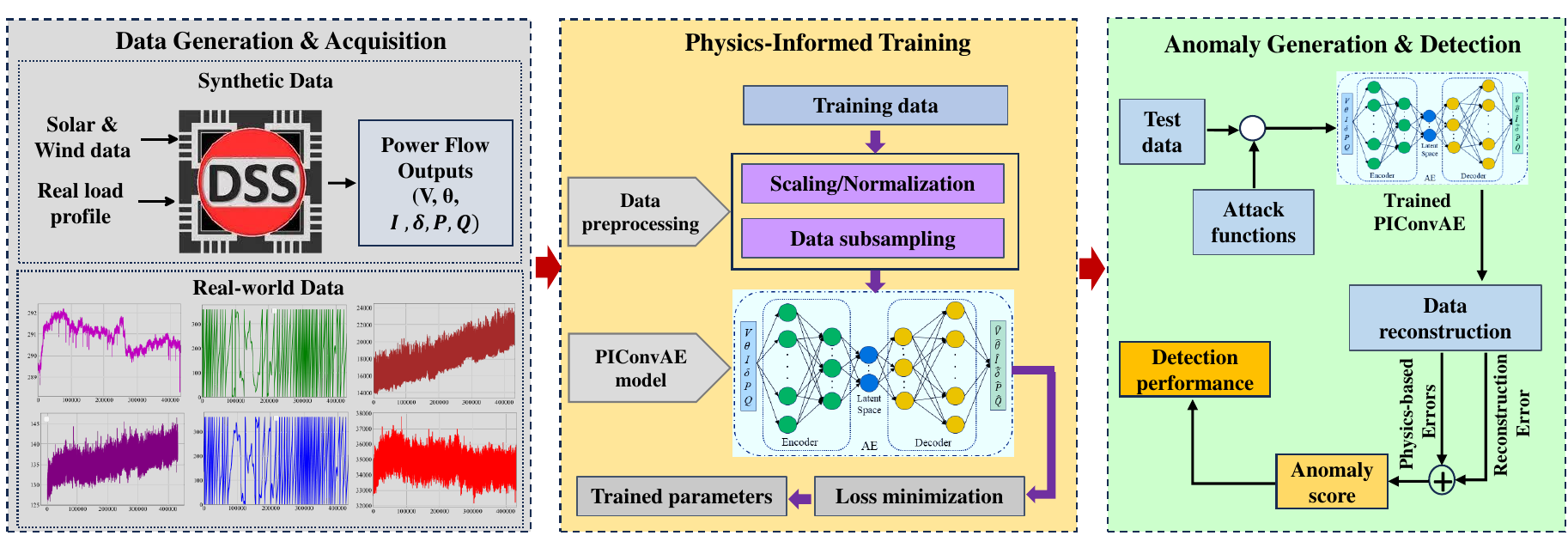}
    \caption{\small The proposed comprehensive framework for anomaly detection}
    \label{fig:methodology}
\end{figure*}
Fig. \ref{fig:methodology} shows the comprehensive framework for anomaly detection including data generation, the training phase of the proposed physics-informed model, and testing procedures. The training data can be either generated using power system power flow software such as the Open distribution system simulator (OpenDSS) or collected through a set of real measurement devices installed in the distribution grids. In the training phase, the real or synthetic measurements are preprocessed and used to train the physics-informed model. Finally, the trained model is employed in the test phase to identify the anomalous points in the manipulated test dataset. In the following, the main concepts and principles of the physics-informed learning model are described and formulated.

\subsection{Problem Setup}
We study the main principles of PINN models and describe how the physical laws can be integrated into the loss function of NN models for improved predictions. Let's assume that the general form of PINN models can be modeled through the following equation
\begin{equation}\label{general_PINNN}
f:= u(t,\theta) + \mathcal{N}(u), \quad t\in [0, T]
\end{equation}
where $u(t,\theta)$ denotes the solution of the NN model with learning parameter $\theta$ at time $t$. The operator $\mathcal N(.)$ represents the nonlinear relationships of the outputs of the data-driven model. The nonlinear operator includes a wide range of nonlinear dynamical and algebraic equations that outline the underlying physical laws in mathematical physics and nonlinear dynamical systems.

\subsection{Underlying Physical Principles}
In the power system domain, generation and demand should be in balance at every timestamp. This means that given the changes in the generation of traditional generators and DERs, and load fluctuations, the power flow requirements need to be always met. Measurement devices installed in different buses provide high-resolution real-time power flow measurements including voltage and current phasors at each node as well as active and reactive power measurements. Using these measurements, the system conditions such as system stability and any abnormal situations can be identified. The measured signals at each bus should meet the power flow criteria or Kirchhoff's law to indicate the system's normal operational conditions. The Kirchoff's law is expressed as
\begin{align}\label{Kirchhoff's law}
P+jQ &= (V \angle \theta)(I \angle \delta)^* = VI \angle (\theta - \delta)\\
    &=VI cos(\theta - \delta)+jVI\sin(\theta - \delta)
\end{align}
where $V$ and $I$ represent voltage and current magnitudes, $\theta$ and $\delta$ are voltage and current phase angles, and $P$ and $Q$ are active and reactive power measurements at each bus.

\subsection{Neural Network Framework}
In this section, the main framework for the unsupervised neural network model for cyber anomaly detection is formulated.

The common idea for developing an unsupervised learning model for anomaly detection is using the power of reconstruction-based deep learning models. Autoencoder as a class of reconstruction-based models transforms data from high-dimensional space to a latent domain with lower dimensional while preserving the most important features of the training data. The model is trained on a set of normal data to learn the normal behavior of the system by minimizing the following loss function as an objective function;
\vspace{-1mm}
\begin{equation}\label{AE_loss}
\vspace{-1mm}
\mathcal{L_{AE}}\left( \theta, x \right) =\frac{1}{m}\sum_{k=1}^{m}\left| \hat{x}(\theta,x^k)-x^{k} \right|^{2}
\vspace{-1mm}
\end{equation}
This function denotes the mean square error of the input training data $x$ and the reconstructed data $\hat{x}$. $m$ represents the number of training data. The training process of the autoencoder solely relies on the data and does not involve the system's physical principles. Applying the underlying physical knowledge of power systems in (\ref{Kirchhoff's law}) to the data-driven principles of the autoencoder's loss function makes the physics-informed autoencoder model. In this physics-informed learning model, the physical laws are integrated and modeled as regularization terms to constrain the space of admissible solutions.
\begin{figure*}
    \centering
      \includegraphics[clip,trim=0.5cm 3cm 4.4cm 3.3cm, width=0.9\linewidth]{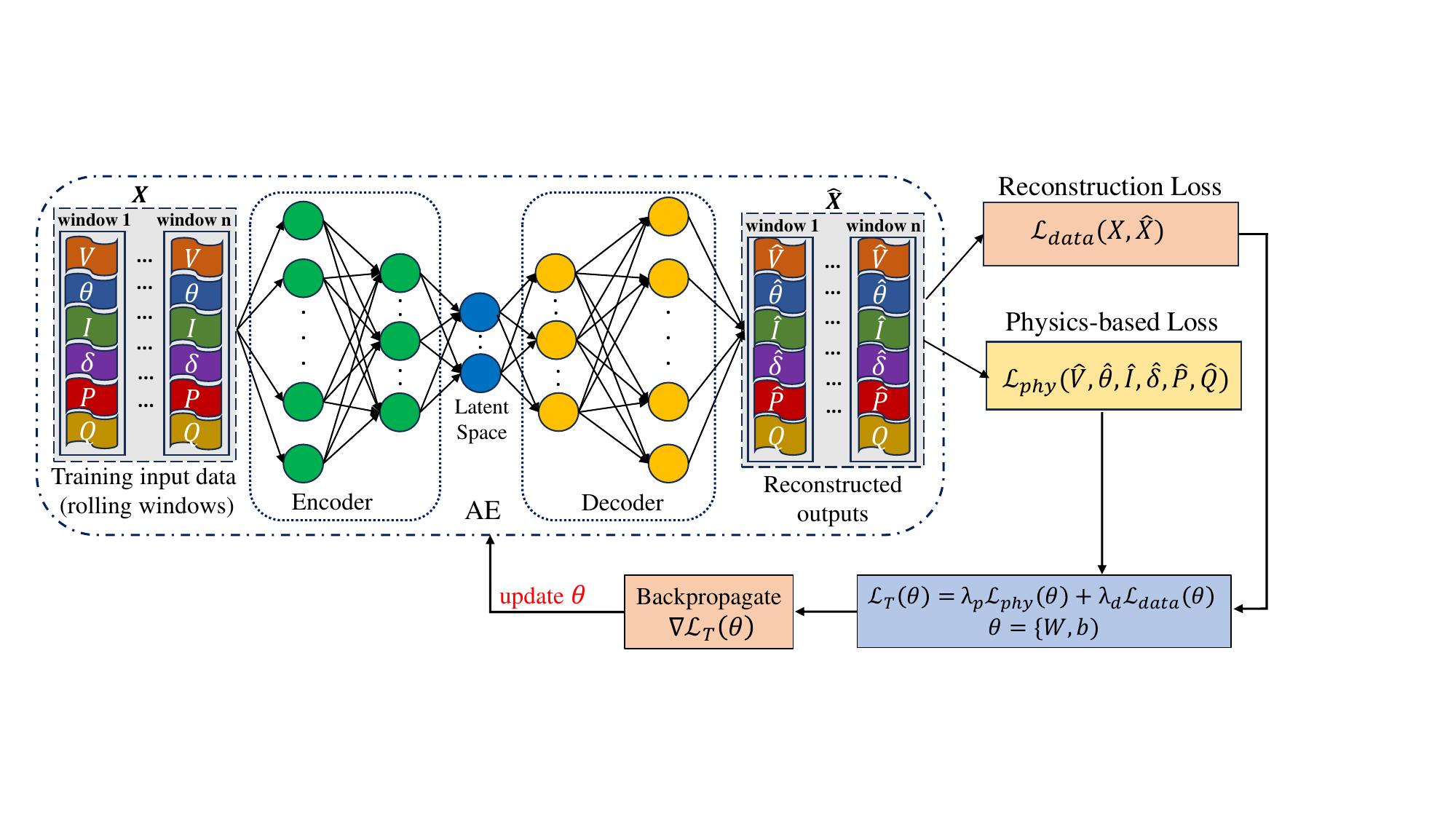}
 \vspace*{-4mm}
 \caption{\small Architecture of physics-informed autoencoder.}
 \label{fig:PIConvAE_framework}
 \vspace*{-4mm}
\end{figure*}
The proposed physics-informed autoencoder model is shown in Fig. \ref{fig:PIConvAE_framework}. The model receives multivariate time-series data $X=[V, I, \theta, \delta, P, Q]$ as input. The goal of the multivariate PIConvAE model is to minimize the following loss and reconstruct the data to be as similar as possible to the original training data.
\begin{equation}\label{PIConvAE_loss}
\mathcal{L}_{T}(\theta, X, \hat{X})= \alpha_d \mathcal{L}_{data} + \alpha_{phy}( \mathcal{L}_{phy_P} + \mathcal{L}_{phy_Q})
\end{equation}
where $\hat{X}$ represents the reconstructed data vector or the solution of the neural network model, $\alpha_d$ and $\alpha_{phy}$ are used to balance the contributions of data-driven and physics-informed losses. The $\mathcal{L}_{data}$ is the modified loss function of (\ref{AE_loss}) for a multivariate input data defined as
\begin{equation}\label{multivariate_AE_loss}
\vspace{-3mm}
\mathcal{L}_{data}\left( \theta, x \right) =\frac{1}{n \times m}\sum_{l=1}^{n} \sum_{k=1}^{m}\left| \hat{x}(\theta,x_l^k)-x_l^{k} \right|^{2}
\vspace{2mm}
\end{equation}
where $x \in X$ and $\hat{x} \in \hat{X}$, $n$ is the number of input features for training ($n=6$). The physics-based loss functions $\mathcal{L}_{phy_P}$ and $\mathcal{L}_{phy_Q}$ are defined as follows.
\vspace{-1mm}
\begin{equation}\label{PI_loss_P}
\mathcal{L}_{Phy_P}\left(\hat{x} \right) =\frac{1}{m}\sum_{k=1}^{m}\left| \hat{P}^k-\hat{V}^k \hat{I}^k \cos{(\hat{\theta}^k-\hat{\delta}^k)} \right|^{2}
\end{equation}
\vspace{-1mm}
\begin{equation}\label{PI_loss_Q}
\mathcal{L}_{Phy_Q}\left(\hat{x} \right) =\frac{1}{m}\sum_{k=1}^{m}\left| \hat{Q}^k-\hat{V}^k \hat{I}^k \sin{(\hat{\theta}^k-\hat{\delta}^k)} \right|^{2}
\vspace{-1mm}
\end{equation}
These physics-aware loss functions force the reconstructed time-series measurements to follow Kirchhoff's law and penalize any outputs that deviate from physically consistent solutions. The choice of data-driven and physics-based coefficients is an open area and their values are found based on trial and error. We have found that the best choice for better predictions is where their effects are the same, i.e., $\alpha_{d}=\alpha_{phy}=1$.


\section{Experimental Results} \label{Results}
\subsection{Test Systems and Data Description}
\begin{figure}[t]
    \centering
      \includegraphics[clip,trim=7.4cm -0.3cm 4.9cm -0.53cm, width=1.0\linewidth]{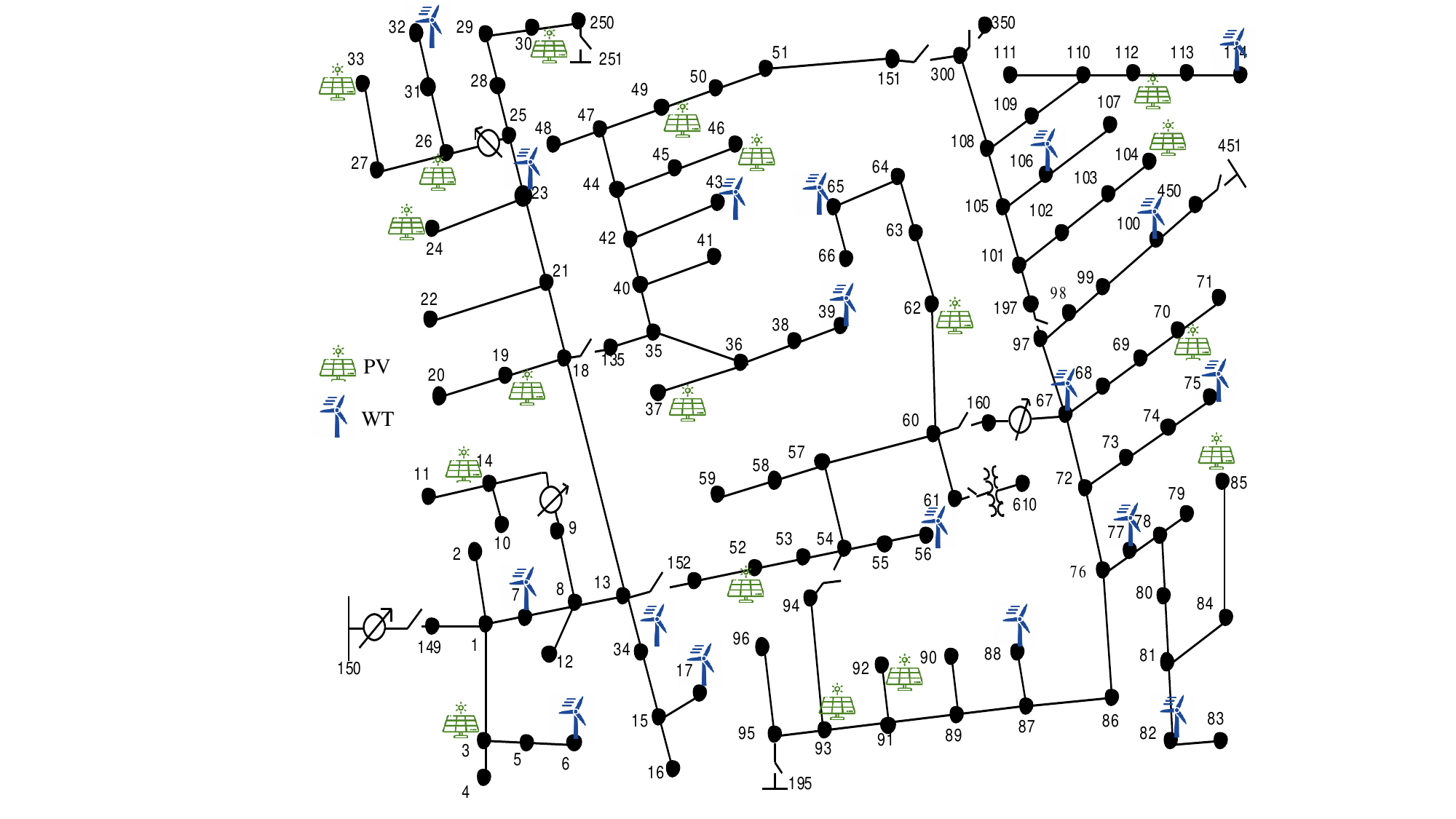}
\caption{\small  Modified IEEE 123-bus system with high penetration of DERs.}
 \label{fig:123_system}
 \vspace{-4mm}
\end{figure}

In this section, simulations are performed on the modified 123-bus test system and real-world $\mu$PMU data from a feeder in Riverside, CA.
Fig. \ref{fig:123_system} shows the modified IEEE 123-bus integrated with high penetration of renewable energy sources, i.e., PV solar panels and wind turbine generation. To integrate the uncertainty and randomness of DER generations and load fluctuations into the 123-bus system, real-world data of solar irradiance, wind speed, and ambient temperature of the city of San Diego in the first week of 2021 is employed \cite{zideh2024unsupervised}. The power flow simulation results including node voltage and current phasors, and node and line active and reactive powers are generated by implementing the system in OpenDSS software for a week. 70\% of 10080 generated data points are used for training, 10\% for validation, and 20\% for testing.
\begin{figure*}
    \hspace{0mm}
    \centering
    \includegraphics[clip,trim=1.3cm 20.5cm 1.45cm 2.2cm, width=0.99\textwidth]{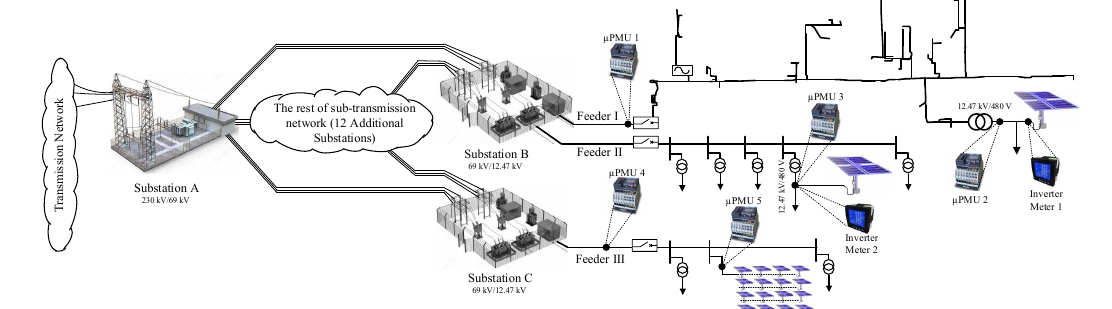}
    \caption{\small Test system with real-world feeders in Riverside, CA \cite{shahsavari2017autopsy}.}
    \label{fig:RCA_system}
\end{figure*}
The second system to perform the simulation studies is the real-world distribution feeder installed in Riverside, CA \cite{shahsavari2019situational, aligholian2021unsupervised, mohsenian2022smart}. Fig. \ref{fig:RCA_system} illustrates the real-world test case where the data from $\mu$PMU 2 installed in the 480-Volt load side of Feeder I is utilized for evaluating the PIConvAE model. The sampling rate of $\mu$PMUs installed in RCA is  120Hz meaning that they collect 120 measurements for each second. With this sampling rate, the collective measurements for each hour are 432,000. We use the time-series measurements including voltage and current phasors, and active and reactive powers of one hour to carry out the simulations.

\subsection{Data Preprocessing}\label{parameters}
The input training data might include missing data. This requires preprocessing the training data before feeding them to the PIConvAE model. Additionally, the training data should be normalized to control the model's sensitivity to very small or large values. Normalization also brings all the training features to the same scale and makes them equally important for the model in the training process. The min-max scaling strategy is utilized in this paper to normalize the input features between -1 and 1. Moreover, due to the temporal relationships in the time-series data, it is not feasible to shuffle them like image-based datasets. To preserve the temporal patterns of the data, the rolling-based strategy is employed where the whole input data with size $m$ is divided into $n$ equally-sized windows ($n=\frac{m-N_w}{N_s}$) with step size $N_s$. This step size is the number of input data each rolling window jumps to get a new window. There should be a trade-off when choosing the window size. Rolling windows with larger window sizes result in higher computational time. Performing different simulations with different window sizes, we noticed that window size 100 and step size 1 provide better training results.

\subsection{Configuration of Training Model}\label{parameters}
The simulation settings of the multivariate PIConvAE model are presented in Table \ref{tab:autoencoder_specifications}. The encoder and decoder each include two Conv1D layers and one Dense layer. The input size is the same as the training window size, i.e., 100. The input data consists of six features each with a dimension of 20 in the latent space making the last layer of the encoder 120. The flattening and reshaping operations are performed for the dense layers to ensure compatibility of data sizes between such layers and previous layers. The slopes of the LeakyReLU functions in the hidden layers are set to 0.2 while the Tanh function is applied to the last layer to limit the outputs between -1 and 1. The learning rate is initially set to 0.001 to explore the space of training parameters and is decayed with a decaying factor of 0.99 in each epoch. As the neural network models tend to overfit in the training process, regularization techniques should be applied. Dropout as a most common regularization technique is used to increase the robustness of the model and make the model less sensitive to specific characteristics and patterns of the training data. Using this strategy, some of the neurons are randomly forced to zero helping the model to perform better with unseen data. The dropout rate is chosen as 0.2 in layers of the autoencoder meaning that 20\% of neurons are set to zero within the layers.

\begin{table}[t]
    \centering
    {\fontfamily{ptm}\selectfont
    \caption{Autoencoder Specifications}
    \label{tab:autoencoder_specifications}
    \begin{tabular}{@{} l c S[table-format=3.0] c c @{}}
        \toprule
        \textbf{Layer} & \textbf{Type} & {\textbf{Kernels/Units}} & {\textbf{Kernel Size}} & \textbf{Activation} \\
        \midrule
        C1 & Encoder & 64 & 5 & LeakyReLU  \\
        C2 & Encoder & 32 & 3 & LeakyReLU \\
        D1  & Encoder & 120 & - & - \\
        C1 & Decoder & 32 & 3 & LeakyReLU  \\
        C2 & Decoder & 64 & 5 & LeakyReLU  \\
        D2  & Decoder & 600 & - & Tanh \\
        \bottomrule
    \end{tabular}
    }
\vspace{-2mm}
\end{table}

\subsection{Performance Metrics}\label{Performance_metrics}
To validate the detection performance of our proposed multivariate PIConvAE model for the synthetic and real-world $\mu$PMU datasets, three anomaly scores are defined for the testing process. The idea is to combine these scores to create one comprehensive score at each time for model evaluation. The first anomaly score is defined as the reconstruction error between the original test data and the reconstructed output of the multivariate PIConvAE model at timestamp $k$ expressed as follows.
\vspace{0mm}
\begin{equation}
\begin{split}
\label{loss_T}
{a}_{r}^{l,k}\left(x^{l,k},\hat{x}^{l,k} \right) &= |x^{l,k}-\hat{x}^{l,k}|
\end{split}
\end{equation}
where $l \in \{V,I,\theta,\delta,P,Q\}$ represents the input features or measurements. Since the anomalies are in the voltage measurements, the reconstruction error for voltage magnitude measurements are calculated . 
The other two scores represent the physics-driven scores that present the deviations of the test data from the physics-based  relationships. The physics-driven anomaly scores for the voltage magnitude measurements at timestamp $k$ are represented as
\vspace{-2mm}
\begin{align}
\label{P_score}
{a}_{p}^{V,k} &= |V^k-\frac{P^k}{I^k \cos(\theta^k-\delta^k)}|
\end{align}
\vspace{-6mm}
\begin{align}
\label{Q_score}
{a}_{q}^{V,k} &= |V^k-\frac{Q^k}{I^k \sin(\theta^k-\delta^k)}|
\vspace{+1mm}
\end{align}
The combination of these three scores is defined as follows to be used for performance evaluation.
\vspace{0mm}
\begin{equation}
\begin{split}
\label{comp_score}
{a}^{V,k}\left(x^{V,k},\hat{x}^{V,k} \right) &= {a}_{r}^{V,k} + {a}_{p}^{V,k} + {a}_{q}^{V,k}
\end{split}
\end{equation}

\begin{table}[htbp]
\centering
{\fontfamily{ptm}\selectfont
\caption{Formulas for performance metrics}
\label{metrics_formula}
\begin{tabular}{| >{\centering\arraybackslash}m{1.3cm} | >{\centering\arraybackslash}m{2.1cm} | >{\centering\arraybackslash}m{1.3cm} | >{\centering\arraybackslash}m{2.1cm} |}
\hline
\textbf{Metric} & \textbf{Formula} & \textbf{Metric} & \textbf{Formula} \\ 
\hline
\textbf{Accuracy}& \vspace{1mm} $\frac{TP + TN}{TP + TN + FP + FN}$ \vspace{1mm}& \textbf{Recall} & \vspace{1mm} $\frac{TP}{TP + FN}$ \vspace{1mm}\\
\hline
\textbf{Precision} & \vspace{1mm}$\frac{TP}{TP + FP}$ \vspace{1mm}& \textbf{F-1 Score} & $2 \times \frac{\text{Precision} \times \text{Recall}}{\text{Precision} + \text{Recall}}$\\
\hline
\end{tabular}
}
\end{table}
The calculated anomaly score is utilized to predict the measurements as normal or anomalous. The performance metrics formula is provided in Table \ref{metrics_formula} where TP, TN, FP, and FN represent true positive, true negative, false positive, and false negative, respectively.

\subsection{Anomaly Detection Results}
The performance of the multivariate PIConvAE model is evaluated by training the model on the modified IEEE 123-bus system and the real-world feeder of RCA datasets. The PIConvAE model is trained for 800 epochs on the normal dataset for both systems. After the training process, the trained parameters are employed to test the model by reconstructing the test data and finding the anomaly scores. Furthermore, The threshold is applied based on the three-sigma rule to identify anomalies in the test datasets. 

\begin{figure}[t]
    \centering
      \includegraphics[clip,trim=0.5cm -0.3cm 0.0cm -0.53cm, width=0.98\linewidth]{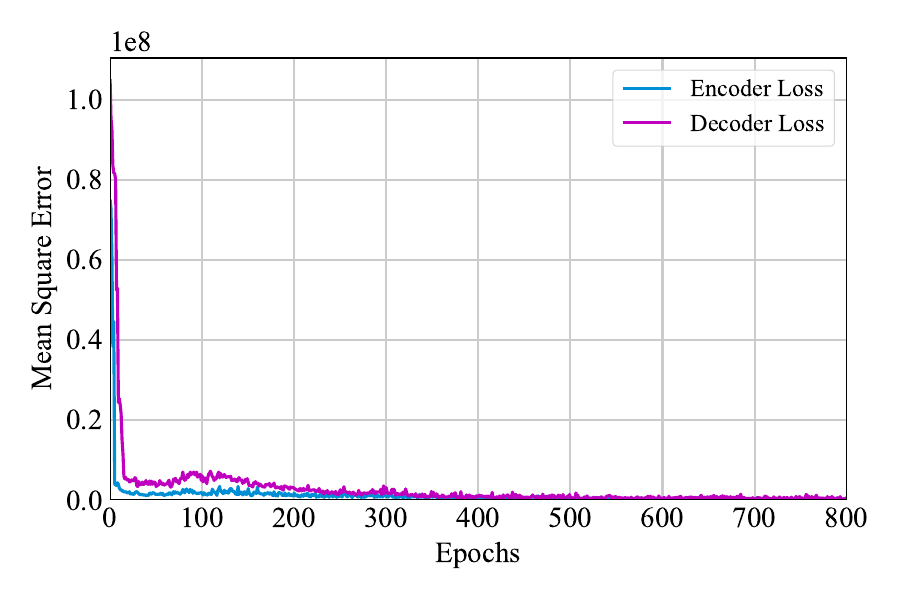}
      \vspace{-2mm}
\caption{\small  Training loss of PIConvAE model for IEEE 123-bus dataset.}
\vspace{-8mm}
 \label{fig:training_loss_123}
\end{figure}

For the 123-bus system, the voltage magnitude measurements at bus 35 are employed to train and test the model. Additionally, within the test dataset of the 123-bus system, 105 data points are manipulated using the attack functions described in Section \ref{cyber_Development}.
Fig. \ref{fig:training_loss_123} shows the loss function of the proposed model trained on the 123-bus dataset. The detection performance of the physics-driven model for 123-bus system is presented in Table \ref{tab:prediction_results}. As can be seen from this table, the model achieves a high accuracy rate of 99.21\% demonstrating the high correct predictions and is very effective in distinguishing between anomalies and normal data points. Although the model's recall is lower than other metrics due to some false negatives, with a precision of 98.95\%, it achieves a remarkable performance with very few false positives indicating a high rate of correct anomaly detection. Moreover, as the evaluated dataset is imbalanced for anomaly detection meaning that the portion of normal data is considerably higher than anomalies, the harmonic mean of these two metrics, F-1, is very useful to show the overall performance of the model. The F-1 score of 94\% shows that not only is the model accurate in predicting anomalies but it also can effectively identify the majority of actual anomalous points. 
\vspace{-2mm}
\begin{table}[b]
    \centering
    {\fontfamily{ptm}\selectfont
 \caption{Prediction Results of PIConvAE for two distribution systems}
    \label{tab:prediction_results}
    \begin{tabular}{@{} l S[table-format=2.2] S[table-format=2.2] S[table-format=2.2] S[table-format=2.2] @{}}
        \toprule
        {Metric} & {IEEE 123-bus system (\%)} & {Riverside feeder (\%)} \\
        \midrule
        Accuracy  & 99.21 & 99.94 \\
        Precision & 98.95 & 95.12 \\
        Recall    & 89.52 & 99.14 \\
        F-1 Score & 94.00 & 97.09 \\
        \bottomrule
    \end{tabular}
    }
\end{table}
\vspace{2mm}
\begin{figure}[t]
    \centering
      \includegraphics[clip,trim=0.5cm -0.3cm 0.0cm -0.53cm, width=1.0\linewidth]{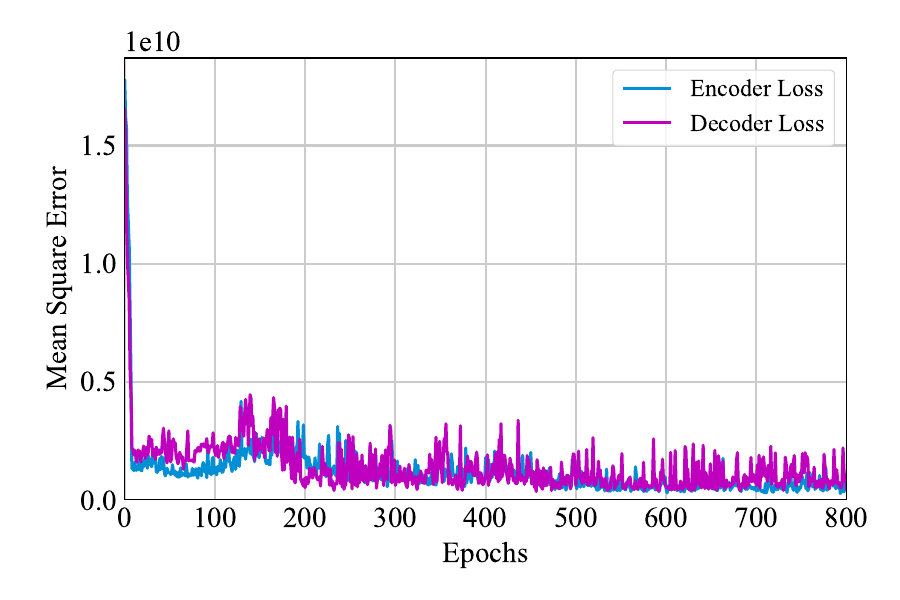}
      \vspace{-6mm}
\caption{\small  Training loss of PIConvAE model for RCA feeder dataset.}
 \label{fig:training_loss_RCA}
 \vspace{-0mm}
\end{figure}

\begin{figure*}[ht]
    \centering
    \begin{subfigure}{0.25\textwidth}
        \includegraphics[width=\linewidth]{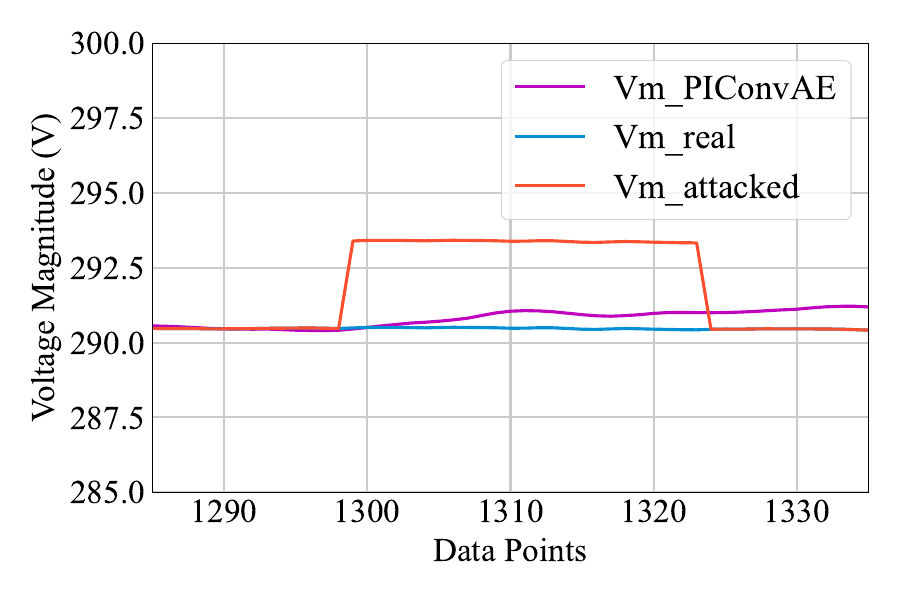}
        \vspace{-5mm}
        \caption{Additive attack}
        \vspace{1mm}
        \label{fig:plot1}
    \end{subfigure}\hfill
    \begin{subfigure}{0.25\textwidth}
        \includegraphics[width=\linewidth]{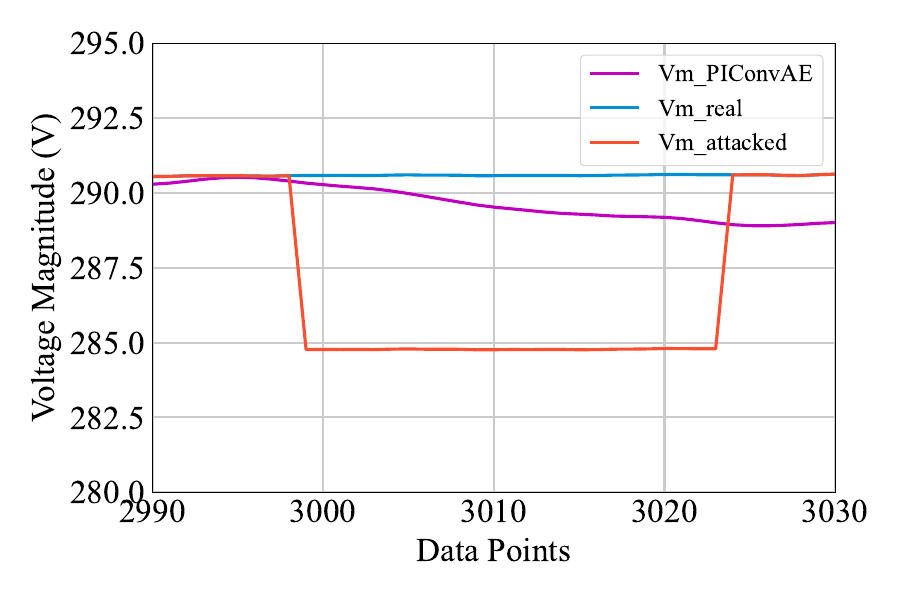}
        \vspace{-5mm}
        \caption{Deductive attack}
        \vspace{1mm}
        \label{fig:plot2}
    \end{subfigure}\hfill
    \begin{subfigure}{0.25\textwidth}
        \includegraphics[width=\linewidth]{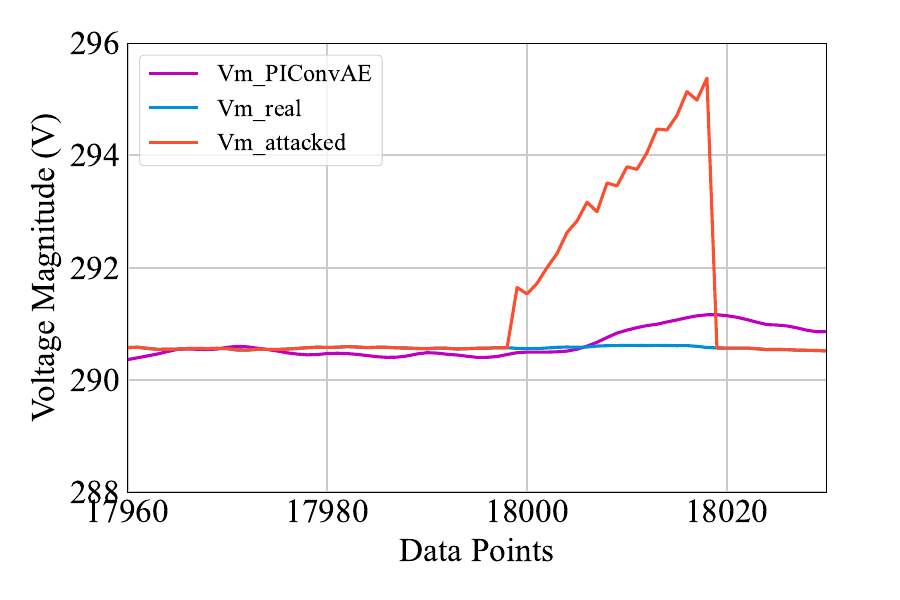}
        \vspace{-5mm}
        \caption{Ramp attack}
        \vspace{1mm}
        \label{fig:plot3}
    \end{subfigure}\hfill
    \begin{subfigure}{0.25\textwidth}
        \includegraphics[width=\linewidth]{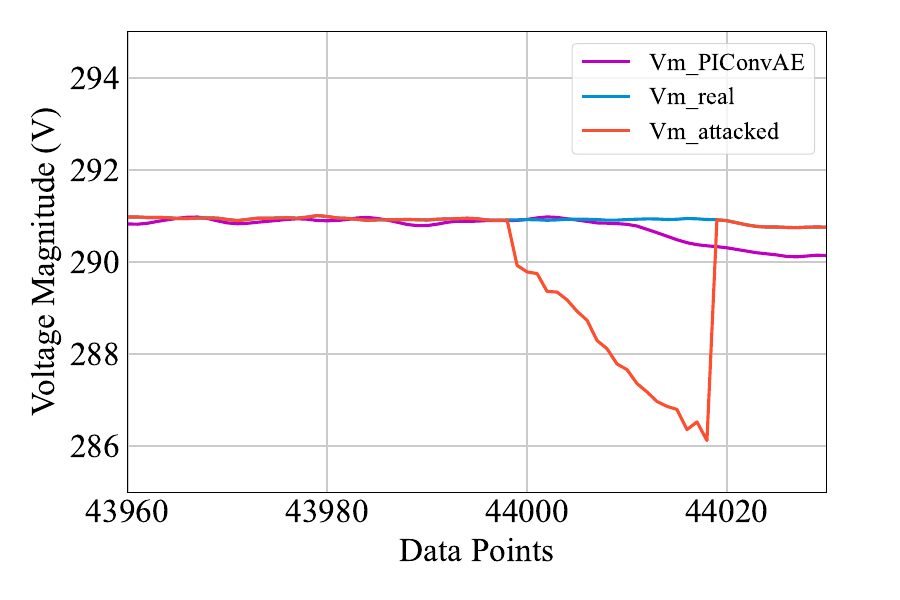}
        \vspace{-5mm}
        \caption{Ramp attack}
        \vspace{1mm}
        \label{fig:plot4}
    \end{subfigure}

    \begin{subfigure}{0.25\textwidth}
        \includegraphics[width=\linewidth]{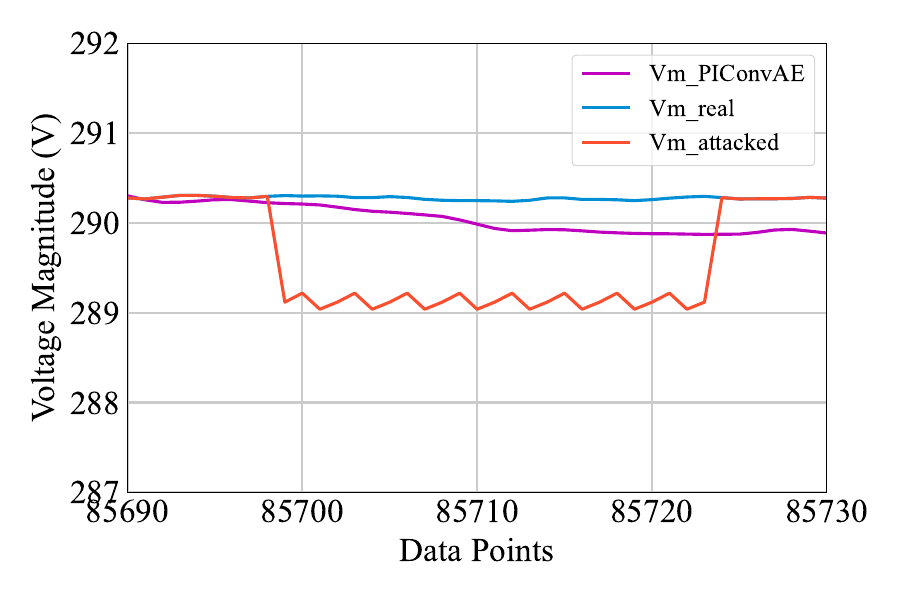}
        \vspace{-5mm}
        \caption{Replay attack}
        \vspace{1mm}
        \label{fig:plot5}
    \end{subfigure}\hfill
    \begin{subfigure}{0.25\textwidth}
        \includegraphics[width=\linewidth]{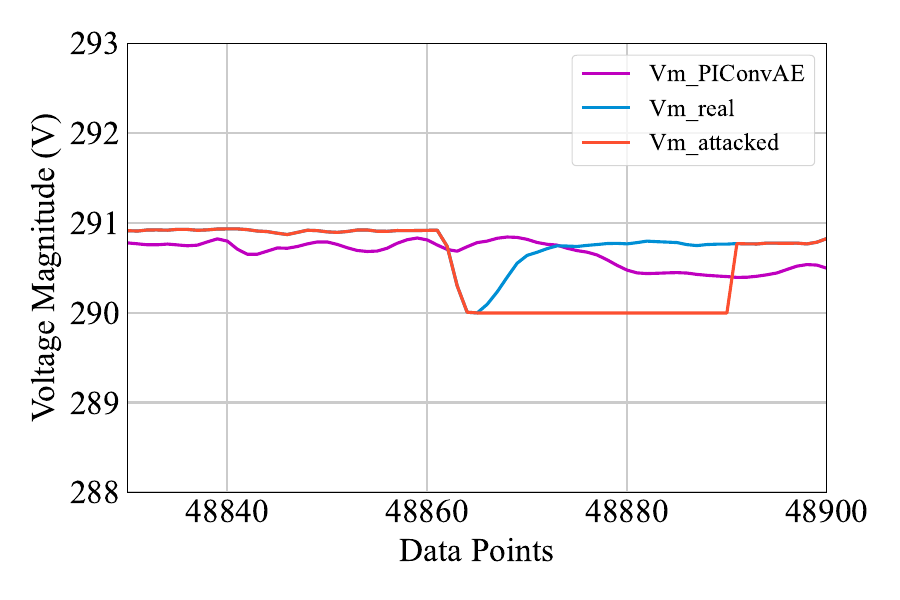}
        \vspace{-5mm}
        \caption{DoS attack}
        \vspace{1mm}
        \label{fig:plot6}
    \end{subfigure}\hfill
    \begin{subfigure}{0.25\textwidth}
        \includegraphics[width=\linewidth]{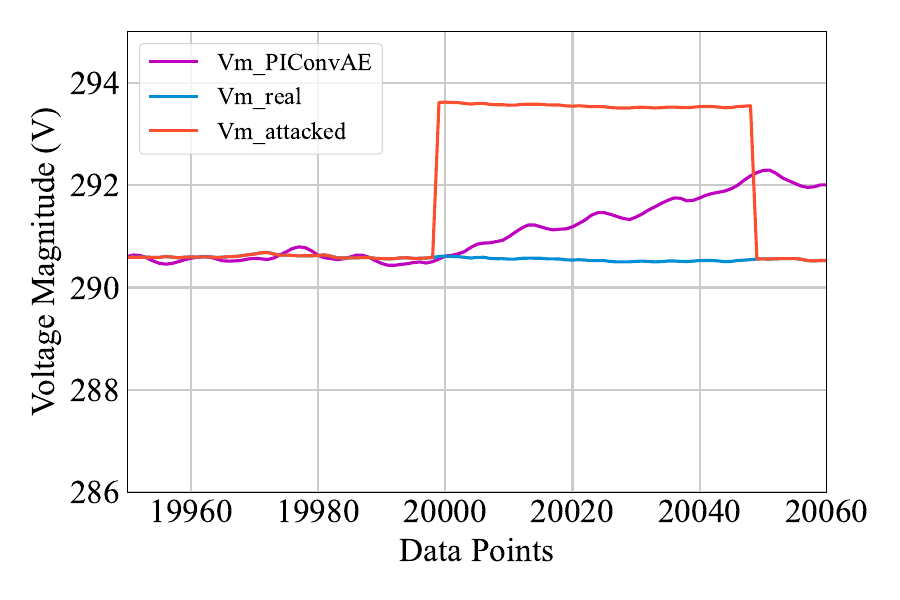}
        \vspace{-5mm}
        \caption{Stealth attack}
        \vspace{1mm}
        \label{fig:plot7}
    \end{subfigure}\hfill
    \begin{subfigure}{0.25\textwidth}
        \includegraphics[width=\linewidth]{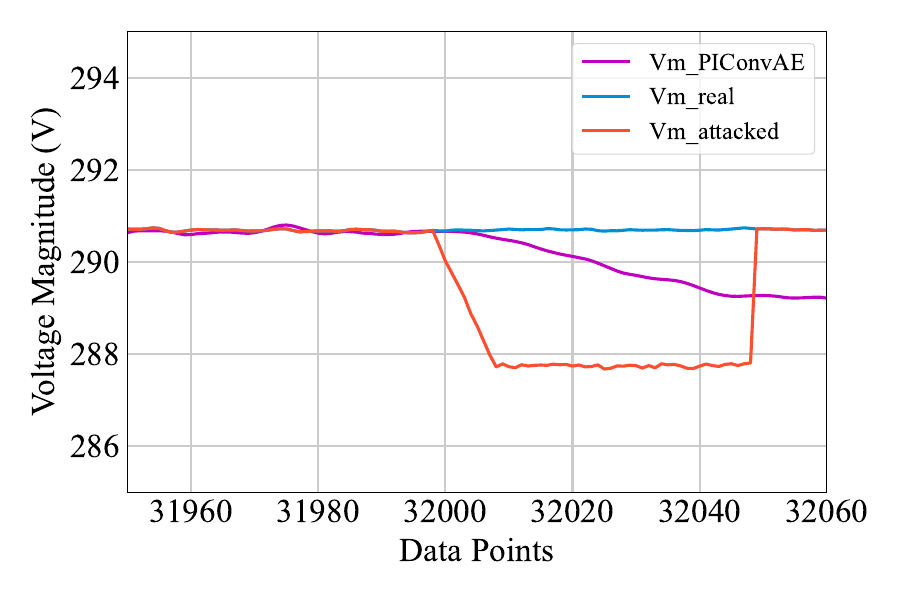}
        \vspace{-5mm}
        \caption{Stealth attack}
        \vspace{1mm}
        \label{fig:plot8}
    \end{subfigure}
    \caption{Comparison of original reconstructed data by PIConvAE Model, test data, and attack data.}
\label{fig:test_data_outputs}
\end{figure*}

The train and test datasets for the real-world feeder in RCA include the voltage measurements of the load side of Feeder I ($\mu$PMU 2 data) in Fig. \ref{fig:RCA_system}. 925 measurements in the test data are manipulated by the anomaly functions for evaluating the performance of the physics-driven model. 
Fig. \ref{fig:training_loss_RCA} shows the loss function of the PIConvAE model trained on one-hour data of the real-world feeder in RCA. 
The model is evaluated using the trained model to identify the cyber anomalies. 
Fig. \ref{fig:test_data_outputs} illustrates the original $\mu$PMU data, manipulated data by attack functions, and the reconstructed data by the PIConvAE model for the test data of the real-world feeder of RCA. As shown in this figure, the PIConvAE model can successfully reconstruct the normal test data across all the attack types. However, the model is unable to fully reconstruct the anomalous data points making a large reconstruction error. This error along with the physics-driven error help the model to precisely identify the attacked data points for all attack types.
The detection performance of the PIConvAE model for the real-world feeder dataset is presented in Table \ref{tab:prediction_results}. As can be seen in this table, the model achieves exceptional performance across all four metrics. The accuracy of 99.94\% indicates that the model can reliably identify anomalous and normal data points with very few overall errors. Additionally, with a high precision of 95.12\% and excellent recall of 99.14, the model can successfully avoid unnecessary false positive alerts and minimize the false negative rate. These high rates of precision and recall lead to a high F-1 score, making the model reliable and effective for real-world applications.

The detection results for both systems demonstrate the reliable and strong performance of the PIConvAE model across both the IEEE 123-bus system and the real-world RCA feeder. This makes the model to be scalable to large-scale systems as well as real-world scenarios with unbalanced configurations. The integration of underlying physical principles into the neural network model enhances the capability of the model to minimize the false positive alerts and risk of undetected security threats.

\subsection{Impact of Training Data Ratio on Detection Results}
\begin{figure}[t]
    \centering
      \includegraphics[clip,trim=0.0cm -0.3cm 0.3cm -0.53cm, width=0.95\linewidth]{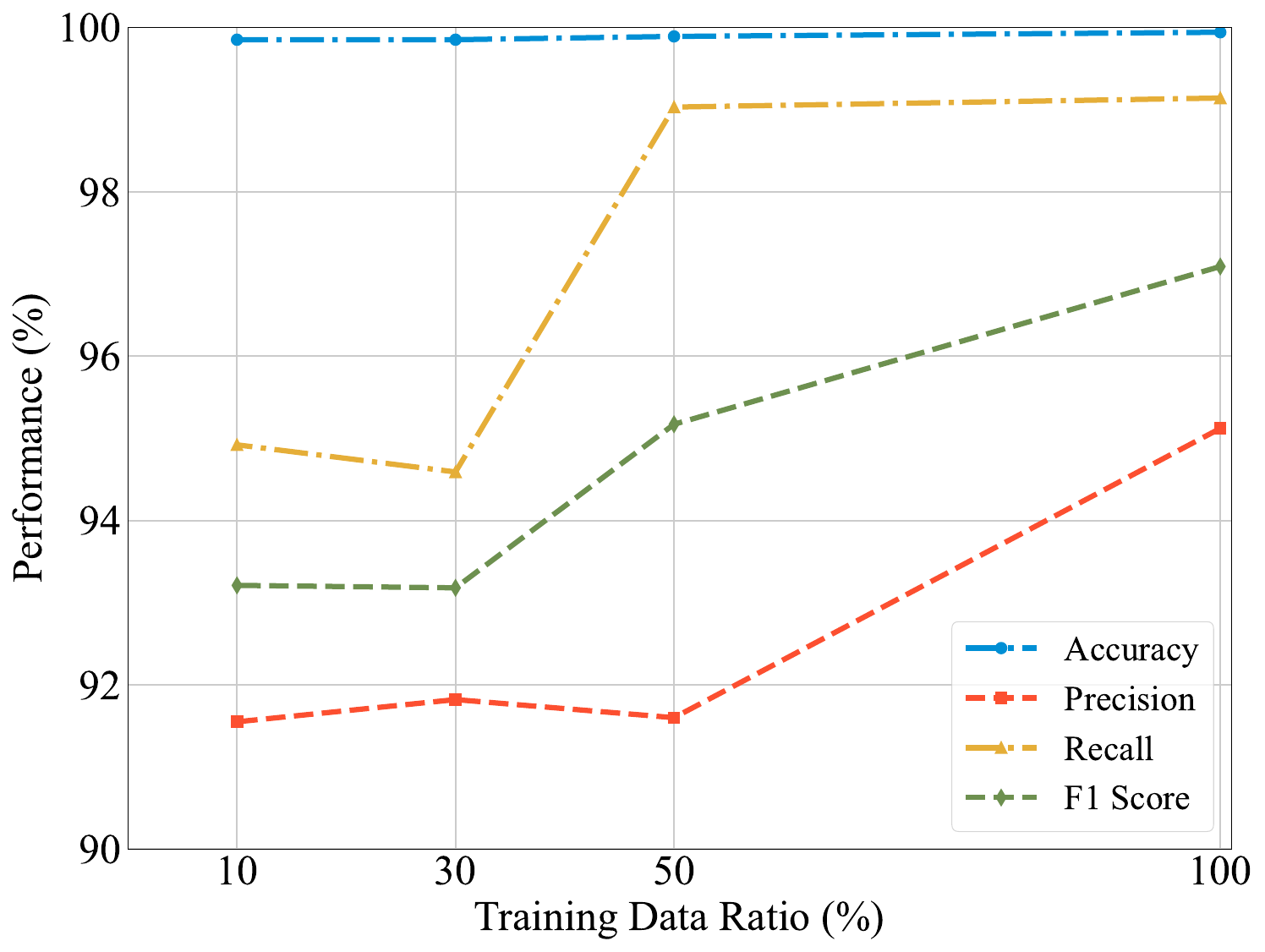}
\caption{\small  Performance of PIConAE model for different training data ratios.}
 \label{fig:data_scarcity}
 \vspace{-3mm}
\end{figure}

One of the main challenges in the training of purely data-driven models is that a vast amount of high-resolution data should be available. In the power system domain with the data scarcity issue, a lower amount of training data degrades the performance of such models. However, physics-informed models have been proposed to overcome this challenge where even with the lack of a high amount of data, their performance is remarkable. In this section, we aim to prove that our proposed multivariate PIConvAE model presents highly accurate detection results with low training data availability. Fig. \ref{fig:data_scarcity} shows the performance of the proposed model with varying training data ratios, i.e., 10\%, 30\%, 50\%, and 100\% for the real-world $\mu$PMU data of RCA. As can be seen from this figure, despite lower training data, the model achieves high accuracy across all data ratios, indicating the strong generalization capability for data scarcity scenarios. Although the model experiences minor fluctuations in both precision and recall, the high values of these metrics with lower data ratios demonstrate its high capability in detecting TPs and minimizing FNs and FPs. This is also shown in the high F-1 score of 95.17\%, 93.18\%, and 93.21\% with training data ratios of 50\%, 30\%, and 10\%, respectively. Overall, the detection results in Fig. \ref{fig:data_scarcity} indicate the significant ability of the proposed physics-informed model to maintain high detection performance under data scarcity scenarios. This shows the robustness of the model to be generalized and well-regularized in real-world scenarios where high-quality data is scarce.

\subsection{Comparison with Existing Machine Learning Models}
The performance of the proposed method is evaluated by conducting a comparative study with data-driven models including convolutional autoencoder (ConvAE), adversarial autoencoder (AAE) \cite{zideh2024unsupervised}, random forest (RF), K-Means, and OneClassSVM (OCSVM). The simulation settings for each of these models are selected such that the maximum detection results are achieved. The descriptions of the data-driven models are provided as follows.\\
\textbf{1) \textit{ConvAE}}: This model is the purely data-driven version of the proposed PIConvAE model where the autoencoder is trained to minimize the reconstruction-based loss function. The model is trained with the same specifications as the physics-informed model to make a fair comparison of these two models. This provides insights into how embedding the underlying physical principles can significantly improve the performance of the PIConvAE model for anomaly detection.\\
\textbf{2) \textit{Adversarial Autoencoder (AAE)}}: This model consists of an autoencoder in the generator network of the generative adversarial networks. The generator's goal is to generate new data to be as similar as possible to the training input data. On the other hand, the discriminator tries to distinguish the generated data of the generator and the real data. The structure and simulation settings of the AAE model are described in \cite{zideh2024unsupervised}. For the data subsampling, the rolling window strategy with a window size of 100 and step size of 1 is applied.\\
\textbf{3) \textit{Random Forest}}: RF is a powerful and popular ensemble machine learning method that creates multiple decision trees and combines their predictions to achieve improved accuracy for detection purposes. This model splits the original dataset into several different datasets and utilizes each of them for training a different decision tree. Since the results of different trees are aggregated, RF is robust to overfitting and generalizes well.\\
\textbf{4) \textit{K-Means}}: K-Means is one of the most popular unsupervised clustering-based machine learning algorithms that is used to partition the data into K clusters. Initially, a centroid is assigned to each cluster and the distances of each datapoint to the clusters' centroid are updated. Eventually, each point is assigned to the cluster with the nearest centroid. 
\\
\textbf{5) \textit{OneClassSVM (OCSVM)}}: Unlike traditional SVM algorithm that is mostly used for multi-class classification, OCSVM is utilized for binary classification to distinguish outlier or anomalies from normal points. As a class of unsupervised learning models, this model is trained on normal datasets and the goal is to find a decision boundary or hyperplane that separates the data points from the origin with the maximum margin. The data points on one side of the hyperplane are considered as anomalies whereas the other side specifies the normal points. 

\begin{figure}[t]
    \centering
      \includegraphics[clip,trim=0.2cm -0.3cm 0.2cm -0.53cm, width=1.0\linewidth]{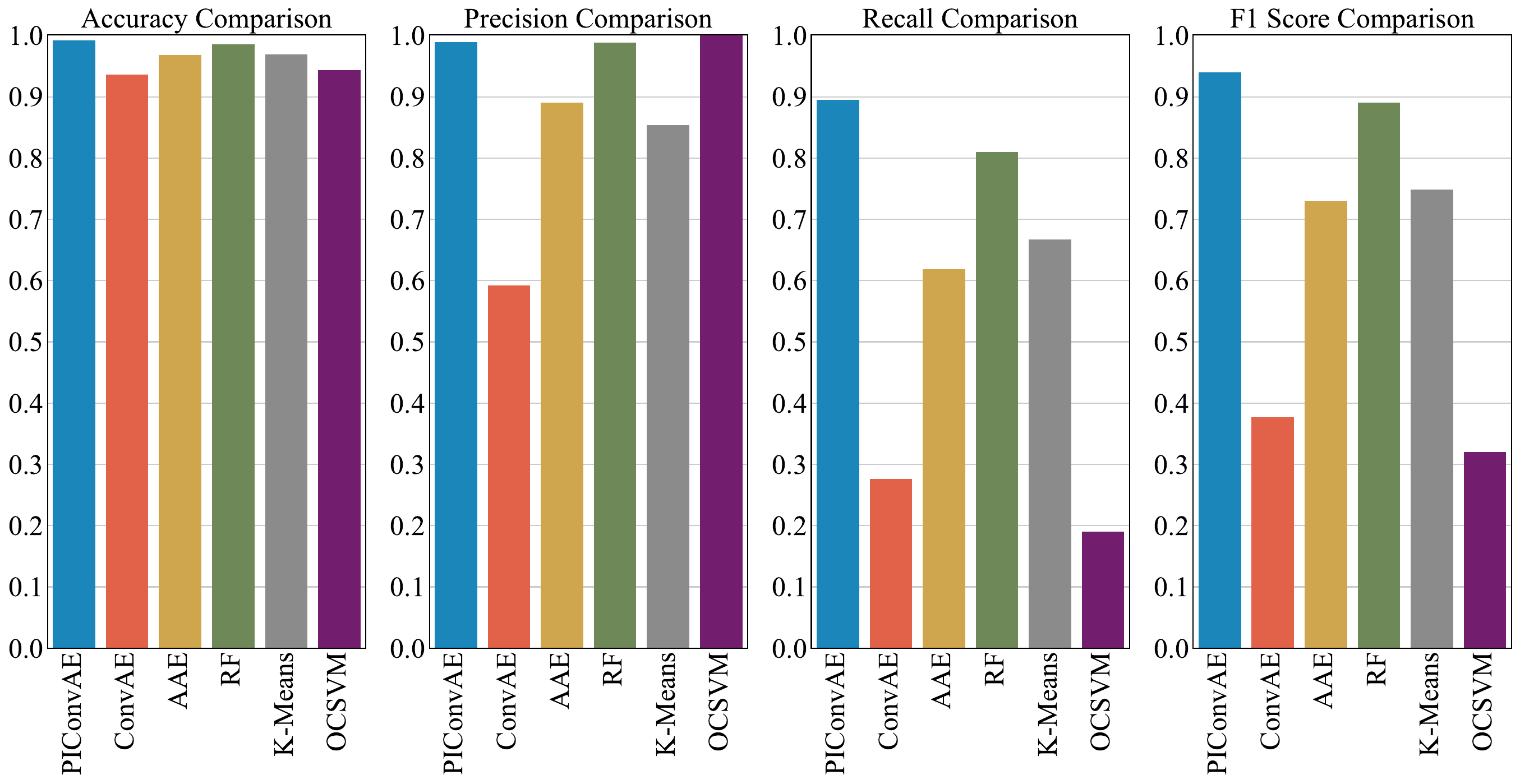}
\caption{\small  Performance comparison for IEEE 123-bus system.}
 \label{fig:models_comparison_123}
 \vspace{-3mm}
\end{figure}

\begin{figure}[t] 
    \centering
    \begin{subfigure}[b]{\columnwidth}
        \centering
        \includegraphics[width=\columnwidth]{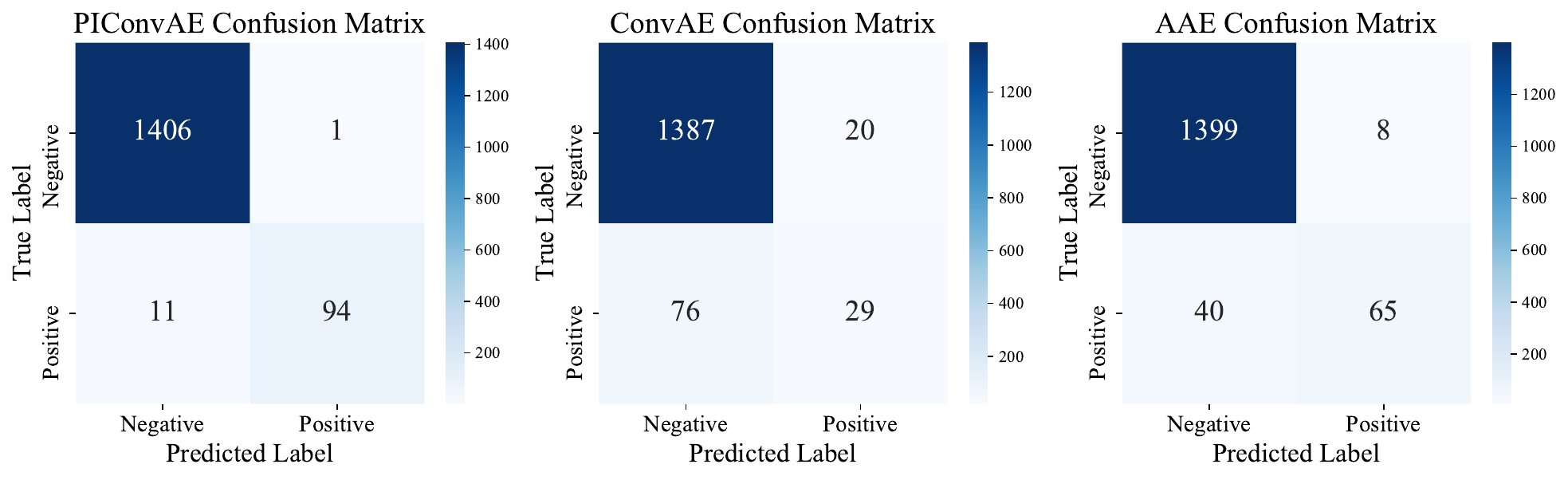} 
    \end{subfigure}
    \vspace{0.5cm} 
    \begin{subfigure}[b]{\columnwidth}
        \centering
        \includegraphics[width=\columnwidth]{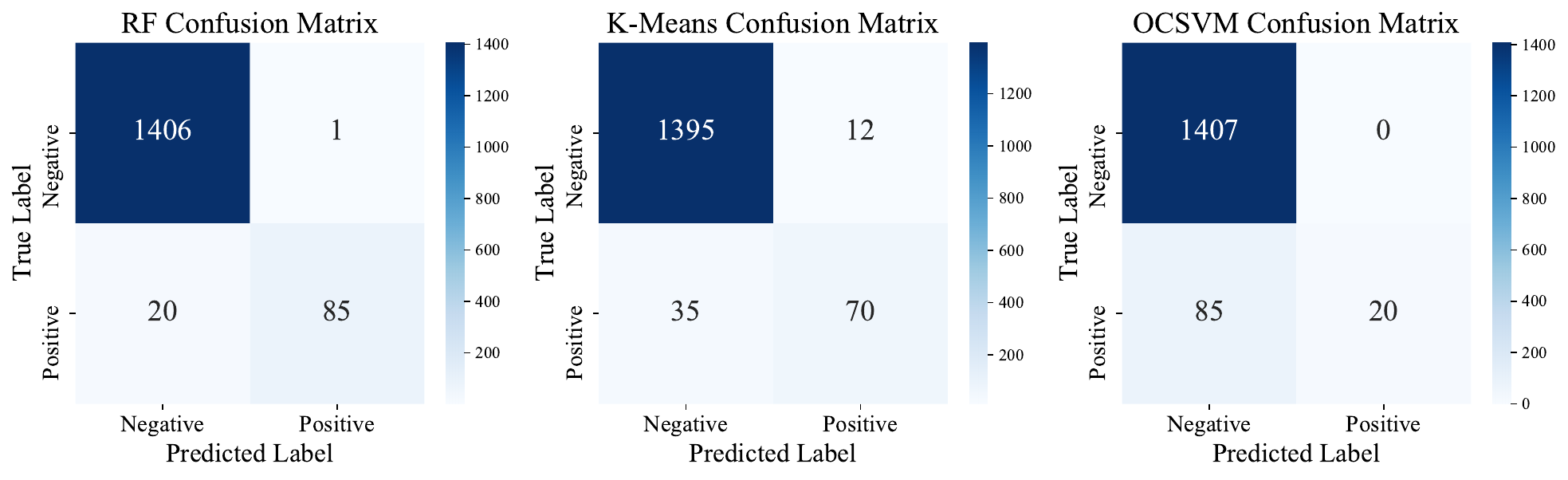} 
    \end{subfigure}
    \vspace{-8mm}
    \caption{Confusion matrices for IEEE 123-bus system.}
    \label{fig:confusion_123}
\end{figure}

Figs. \ref{fig:models_comparison_123} and \ref{fig:confusion_123} present comparative results for the detection performance of PIConvAE and data-driven methods as well as confusion matrices for these models. According to the performance metrics in Fig. \ref{fig:models_comparison_123}, The PIConvAe model achieves the highest accuracy, surpassing the other top detectors such as RF and K-Means with accuracies of 98.61\% and 96.89\%, respectively. The noticeable outcome is that the integration of physical laws into the ConvAE model increases its accuracy from 93.65\% to 99.21\%. In terms of precision and recall, although the OCSVM model outperforms other models with 100\% precision, the significantly lower recall rate, 19.05\%, makes it undesirable to identify anomalies as it misses many anomalous points. The PIConvAE model achieves the second-highest precision and leads in recall to have the highest F-1 score compared to other models. Comparing the detection performance of the ConvAE model with the physics-informed model concludes that the data-driven model experiences a boost in precision and recall and their harmonic mean (F-1 score). The integration of physical laws increases the F-1 score of the ConvAE model from 37.66\% to 94\%. These results are verified by comparing the confusion matrices of the models where the PIConvAE model outperforms other models with the highest true positive (94) and lowest false negative (11). Although RF and OCSVM have very low false positives, they cannot correctly identify the majority of anomalous points.

\begin{figure}[t]
    \centering
      \includegraphics[clip,trim=0.2cm -0.3cm 0.2cm -0.53cm, width=1.0\linewidth]{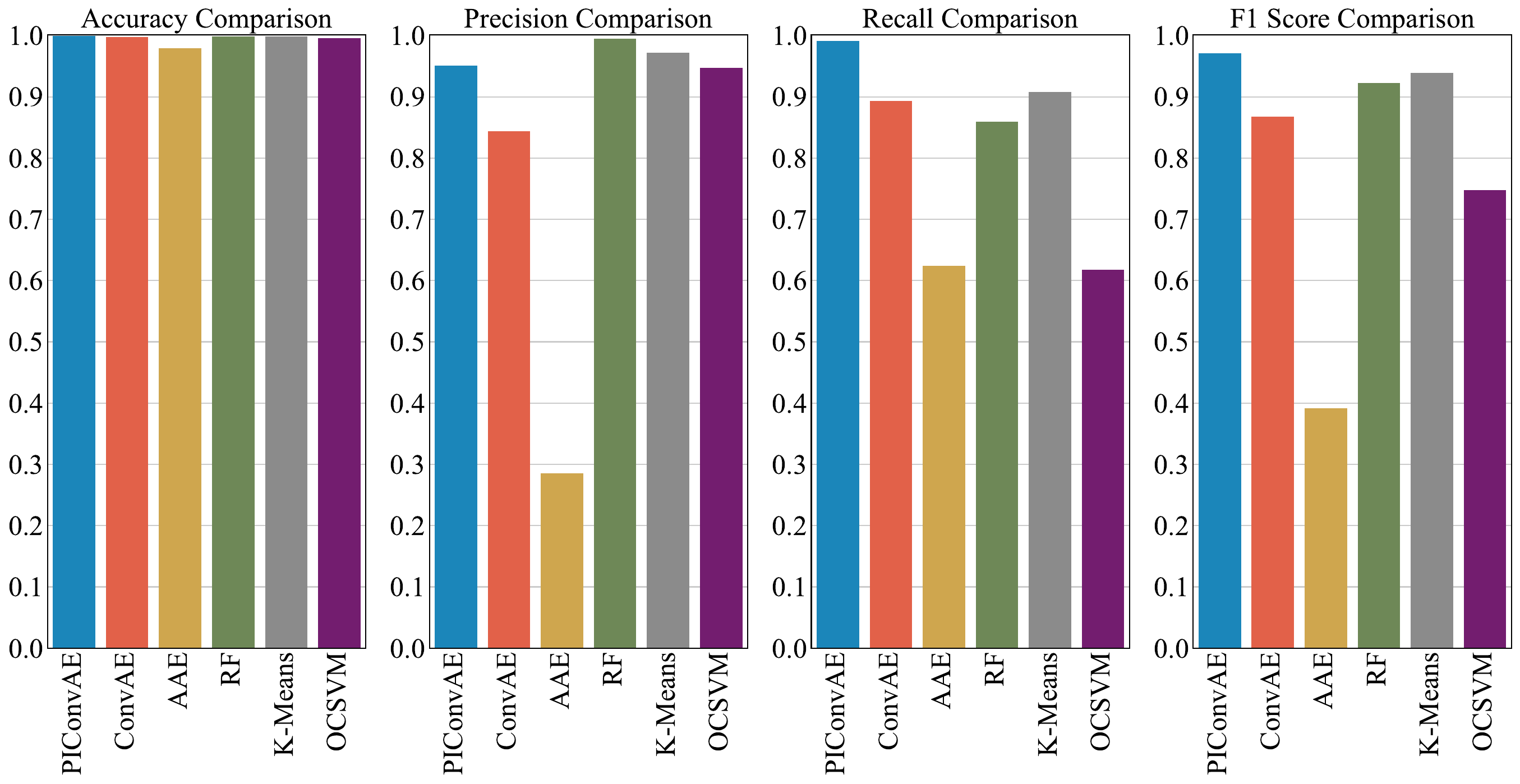}
\caption{\small  Performance comparison for RCA real-world feeder.}
 \label{fig:models_comparison_RCA}
 \vspace{-3mm}
\end{figure}

\begin{figure}[!t] 
    \centering
    \begin{subfigure}[b]{\columnwidth}
        \centering
        \includegraphics[width=\columnwidth]{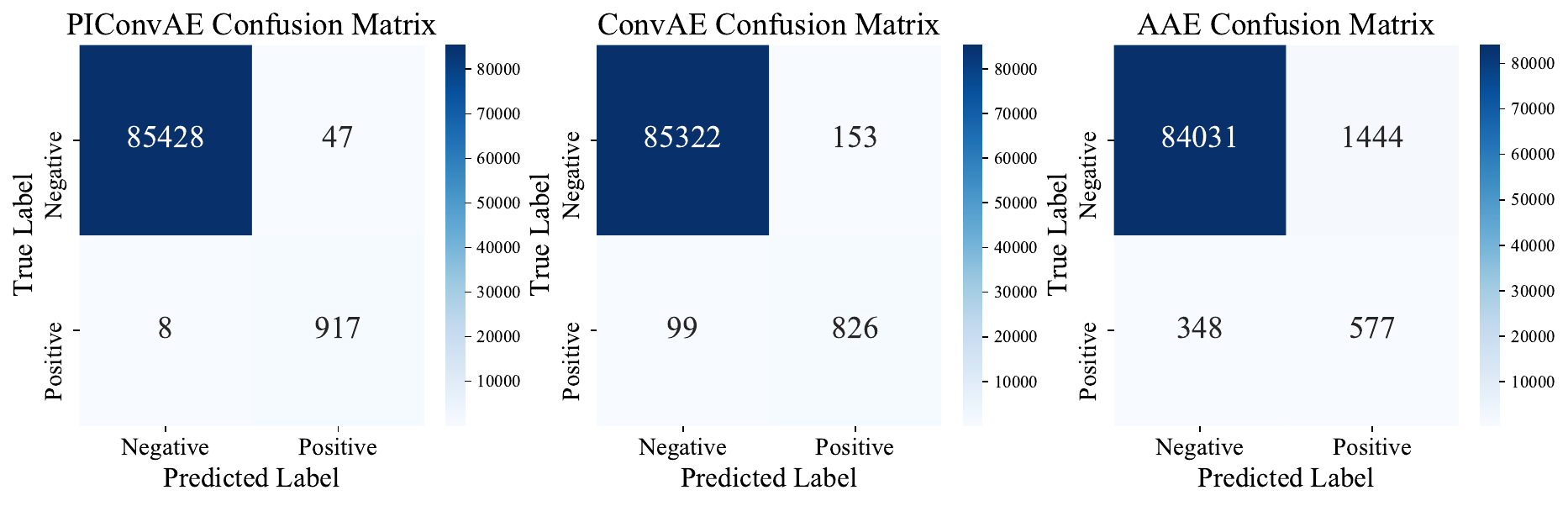} 
    \end{subfigure}
    \vspace{0.7cm} 
    \begin{subfigure}[b]{\columnwidth}
        \centering
        \includegraphics[width=\columnwidth]{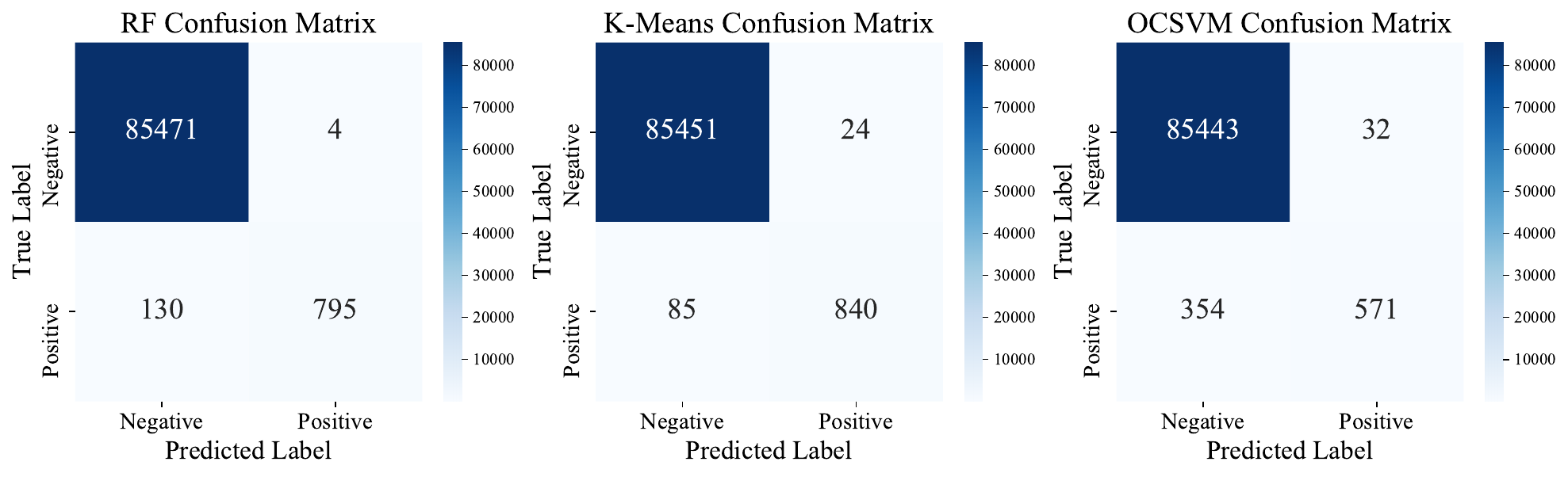} 
    \end{subfigure}
    \vspace{-8mm}
    \caption{Confusion matrices for RCA system.}
    \label{fig:confusion_RCA}
\end{figure}

The comparison of the performance metrics and confusion matrices of the PIConVAE model and other data-driven models for anomaly detection in the feeder of RCA are presented in Figs. \ref{fig:models_comparison_RCA} and \ref{fig:confusion_RCA}, respectively. All models present high accuracy indicating their strong overall performance in the RCA real-world dataset. In terms of precision, RF and K-Means outperform other models with precision of 99.50\% and 97.22\% whereas the AAE model has a poor performance with precision of 28.55\%. However, the PIConvAE model achieves a remarkable recall with a rate of 99.14\% which is significantly higher than other models. This leads to the highest F-1 score of 97.09\% for the PIConvAE model while the second and third F-1 scores belong to K-Means and RF with the rate of 93.91\% and 92.23\%.
It is worth noting that embedding the physical principles in the training process of the ConvAE model has a great impact on the performance of the model where the F-1 score notably increases from 84.37\% to 95.12
The comparison of the confusion matrices also signifies the superior performance of the proposed PIConvAE model where with 917 true positives, it correctly identifies the majority of anomalies while the other models have lower true positives. Considering the performance metrics and confusion matrices, it is apparent that the PIConvAE model considerably outperforms other models and can be reliably employed for anomaly detection purposes. 
\section{Conclusion}\label{conclusion}
This paper presents a novel physics-informed neural network model that integrates the underlying Kirchhoff's circuit laws into the training process of a multivariate autoencoder for cyber anomaly detection in unbalanced power distribution systems with high DER penetrations. The first principles are integrated into the loss function of the AE model to constrain the reconstructed outputs to be consistent with the physics of the system. Data sub-sampling strategy is applied to the time-series datasets using a rolling window strategy to preserve their temporal patterns in the training and test processes. The experimental results are performed on two unbalanced distribution systems, the modified IEEE 123-bus system and the real-world feeder in RCA, to evaluate the model's effectiveness for large-scale systems and real-world scenarios. The simulation results show that the proposed multivariate PIConvAE model has exceptional performance for cyber anomaly detection in both distribution systems. 
As data scarcity is one of the main challenges in the power system domain, the performance of the PIConvAE model is evaluated under different training data ratios, i.e., 50\%, 30\%, and 10\%. The simulation results show that with high-performance metrics across the data ratios, the proposed model can accurately identify anomalies and has excellent capability to extrapolate and generalize under data scarcity scenarios.
In addition, the performance of the proposed physics-driven model is compared with existing machine learning models. The numerical results demonstrate that the PIConvAE model considerably outperforms other models in terms of accuracy and F-1 score verifying the the strengths of the physical laws integration into the autoencoder model.
Future work includes the development of physics-informed learning models to analyze cyber-physical events in inverter-based resource systems in balanced and unbalanced power systems.




\bibliographystyle{ieeetr}
\bibliography{Reference.bib}


\end{document}